\documentstyle[times,epsf]{mn}
\newcommand{\etal}{{et al}\/.}

\begin{document}
\title[BIMA observations of 3C\,20]{85-GHz BIMA observations of the
double-hotspot radio galaxy 3C\,20}
\author[M.J.~Hardcastle \& L.W. Looney]{M.J.\ Hardcastle$^1$ and
L.W.\ Looney$^2$ \\
$^1$ Department of Physics, University of Bristol, Tyndall Avenue,
Bristol BS8 1TL\\
$^2$ Max-Planck-Institut f\"ur Extraterrestrische Physik, Postfach
1312, 85741 Garching, Germany}
\maketitle
\begin{abstract}
We present 85-GHz observations of the archetypal double-hotspot radio
source 3C\,20 made with the BIMA millimetre array. The resolution of
BIMA allows us to separate the two components of the eastern
hotspot. By comparing the BIMA observations with existing VLA data, we
show that the spectra of the two hotspot components are very similar,
despite the clear differences in their radio structure and their wide
separation. We discuss the implications for models of double hotspot
formation. Weak emission from the lobes of 3C\,20 is detected at 85
GHz, at a level consistent with the predictions of standard spectral
ageing models.
\end{abstract}

\section{Introduction}

In the beam model for powerful extragalactic double radio sources,
hotspots (the bright compact regions at the ends of the source) are
the visible manifestation of a strong shock as the relativistic beam
of energetic particles is suddenly decelerated by interaction with the
external medium surrounding the radio lobes.

This model is challenged by the observation that the lobes of
radio galaxies and quasars very frequently have more than one
hotspot. To account for this, we must either assume that the beam
end-point moves about from place to place in the lobe (the `dentist's
drill' model of Scheuer 1982) or that material flows out from the
initial impact point of the beam to impact elsewhere on the lobe edge
[the `splatter-spot' model of Williams \& Gull (1985) or the
jet-deflection model of Lonsdale \& Barthel 1986]. Both these models
predict that one of the hotspots, the one associated with the first
termination of the jet, should be more compact than the other or
others; it is in fact observed that where jets are explicitly seen to
terminate, they always do so in the most compact, `primary' hotspot
(Laing 1989, Leahy \etal\ 1997, Hardcastle \etal\ 1997). But from
single-wavelength radio observations it is very difficult to
distinguish between different models of multiple hot spot formation.

One area in which different models {\it do} make different predictions
is that of the high-frequency spectra of the secondary (less compact)
hotspots. If these are relics left behind by the motion of the jet, as
in the `dentist's drill' model, then in general we expect shock-driven
particle acceleration to have ceased [though, as pointed out by Cox,
Gull \& Scheuer (1991), it is possible for a disconnected hotspot to
continue to be fed for some time if the disconnection occurs at a
significant distance upstream]. Synchrotron losses will then deplete
the high-energy electrons in the secondary hotspot, so that the
secondary hotspot is expected to have a steeper high-frequency
spectrum than the primary. If secondary hotspots continue to be fed by
outflow from the primary hotspot, then there is still an energy supply
and particle acceleration will continue to operate.

To have the best chance of distinguishing between models using these
spectral differences we must observe at the highest available
frequencies, since the synchrotron lifetime of an electron emitting at
a frequency $\nu$ is proportional to $\nu^{-1/2}$. Observations show
that (primary) hotspots often have a spectral cutoff around the
mm-wave region of the spectrum, perhaps corresponding to a maximum
electron energy attained by the acceleration process. Mm-wave
observations therefore provide one of the best opportunities to test
the different models of multiple hotspot formation.

Only a few multiple-hotspot sources have been observed at millimetre
wavelengths with resolution sufficient to distinguish between the
components of the hotspot complex. Two such sources are the nearby luminous
classical double Cygnus A, 3C\,405 (Wright \& Birkinshaw 1984, Wright
\& Sault 1993) and the peculiar radio galaxy 3C\,123 (Looney \&
Hardcastle 2000), both of which show close hotspot pairs (separated by
a projected distance of about 8 kpc and 5 kpc respectively). In both
these cases the radio-to-mm spectra of the two hotspots were found to
be identical within the errors; there was no evidence for a steeper
spectrum in the secondary hotspot.

In this paper we present 85-GHz observations of a well-known
double-hotspot radio galaxy, 3C\,20. The double hotspot in the E lobe
of this $z=0.174$ object was one of the earliest to be resolved in
radio observations (Jenkins, Pooley \& Riley 1977; Laing 1981),
because of the components' large angular separation (6.5 arcsec,
corresponding to 27 kpc). It has since been well observed with the
NRAO Very Large Array (VLA) at various radio frequencies (e.g.\
Hiltner \etal\ 1994, hereafter H94; Hardcastle \etal\ 1997; Fernini
\etal\ 1997). Its 8.5-GHz radio hotspot structure is shown in Fig.\
\ref{radiomap}. The western hotspot, which lies on what is believed to
be the jet side (Hardcastle \etal\ 1997) and is actually a more
compact hotspot pair (or possibly a single hotspot with an extended
tail) is known to be a mm-wave and optical source (H94, Meisenheimer
\etal\ 1989, Meisenheimer, Yates \& R\"oser 1997) with a spectrum
which cuts off in the optical. But the eastern hotspots were not
detected in the optical (H94, Meisenheimer \etal\ 1997) and until now
they have not been observed at high radio frequencies.

Throughout the paper we use a cosmology with $H_0 = 50$ km s$^{-1}$
Mpc$^{-1}$ and $q_0 = 0$. With this cosmology, 1 arcsecond at the distance
of 3C\,20 corresponds to 3.99 kpc. Spectral index $\alpha$ is defined in the
sense $S \propto \nu^{-\alpha}$, and $\alpha_{\nu_1}^{\nu_2}$ denotes
the two-point spectral index between frequencies of $\nu_1$ and $\nu_2$.

\section{Observations and data analysis}

3C\,20 was observed in the B and C configurations of the 10-element BIMA
array\footnote{The BIMA array is operated by the Berkeley Illinois
Maryland Association under funding from the U.S. National Science
Foundation.}. The data were taken on 1999 Oct 11
(C array) and 1999 Oct 30 (B array), with the correlator configured to
give two 800-MHz bands centred on 83.15 and 86.60 GHz. These were
combined in the final images to give an effective frequency of 84.87
GHz. Combining the B- and C-configuration data gives good coverage of
the {\it uv} plane between 2.2 and 66 k$\lambda$, so the resulting
images have a resolution of about 3 arcsec and are sensitive to
structures up to $\sim 50$ arcsec in size, well matched to the largest
angular size of 3C\,20.

The data were processed using the MIRIAD package (Sault, Teuben, \&
Wright 1995). Amplitudes were calibrated using observations of Uranus
(B-array) and Mars (C-array) to bootstrap the flux density of the
phase calibrator, the quasar 0136+478, which was set to 3.68 Jy. The
uncertainty in the amplitude calibration is estimated to be about 7
per cent.

\begin{figure*}
\caption{85-GHz images of 3C\,20. In both images, the contours are at
$(-2, -1, 1, 2, 3\dots 10, 15, 20, 25, 30)$ times the $3\sigma$
level. Negative contours are dashed. The cross indicates the position
of the radio core. (a) Upper panel: full-resolution image. The
restoring beam is a 3-arcsec Gaussian. The $3\sigma$ level is 2.88 mJy
beam$^{-1}$. (b) Lower panel: tapered image. The restoring beam is a
6-arcsec Gaussian. The $3\sigma$ level is 2.22 mJy beam$^{-1}$.}
\label{bimapix}
\epsfxsize 14cm
\epsfbox{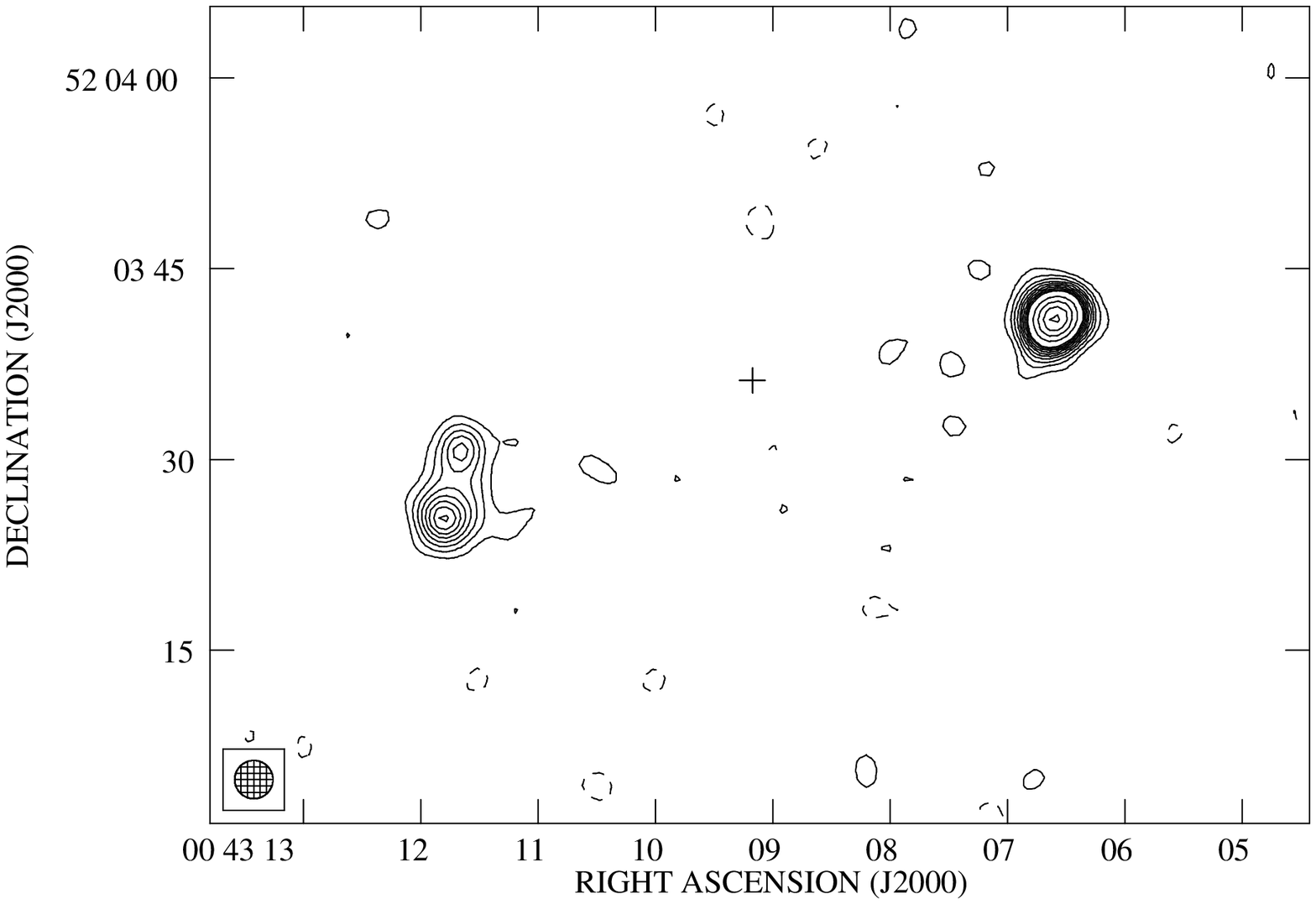}
\epsfxsize 14cm
\epsfbox{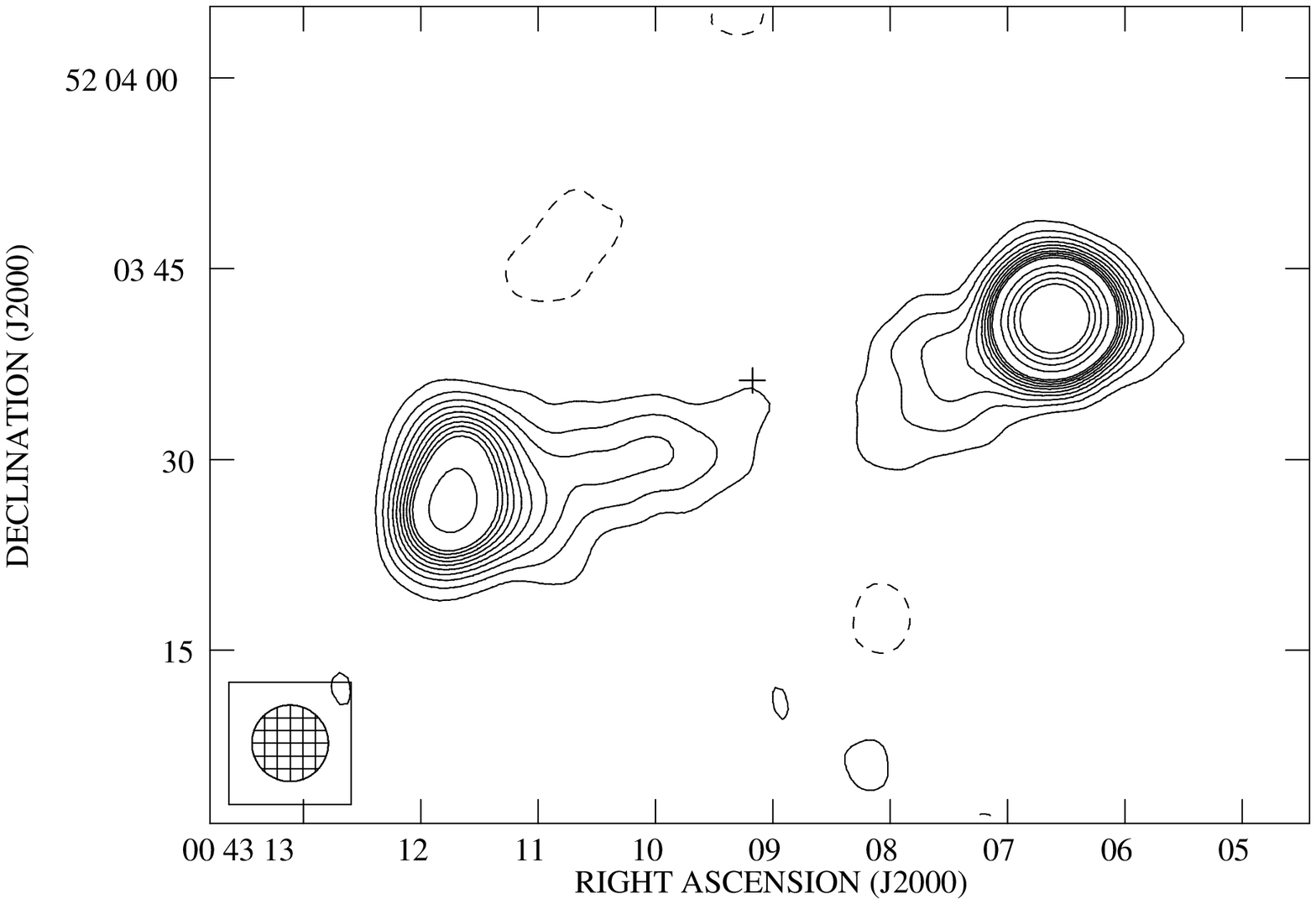}
\end{figure*}

Final images were produced using the AIPS task IMAGR, and are shown in
Fig.\ \ref{bimapix}. The western hotspot and the two eastern hotspots
are clearly detected, and (as shown in Fig.\ \ref{bimapix}b) there is
also some low-surface-brightness emission from both the lobes of
3C\,20.  The weak radio core is not detected at 85 GHz. The net flux density
of the source at 85 GHz is about 0.21 Jy. This is somewhat lower than the
90-GHz flux density of 0.51 Jy quoted by Steppe \etal\ (1988), but the
error on the value of Steppe \etal , though not given explicitly in
their paper, is certainly large.

\section{VLA observations}

\begin{figure*}
\caption{VLA images of 3C\,20 at 1.4, 4.9, 8.5 and 15 GHz. Contours are
at $(-\protect\sqrt 2,-1,1,\protect\sqrt 2, 2, 2\protect\sqrt 2,
4\dots)$ times the $5\sigma$ levels, which are respectively $4.2$,
$0.71$, $0.53$ and $0.78$ mJy beam$^{-1}$. The resolution of all the
images is 3 arcsec.}
\label{vlamaps}
\begin{center}
\leavevmode
\hbox{
\begin{minipage}{8cm}
\epsfxsize 8cm
\epsfbox{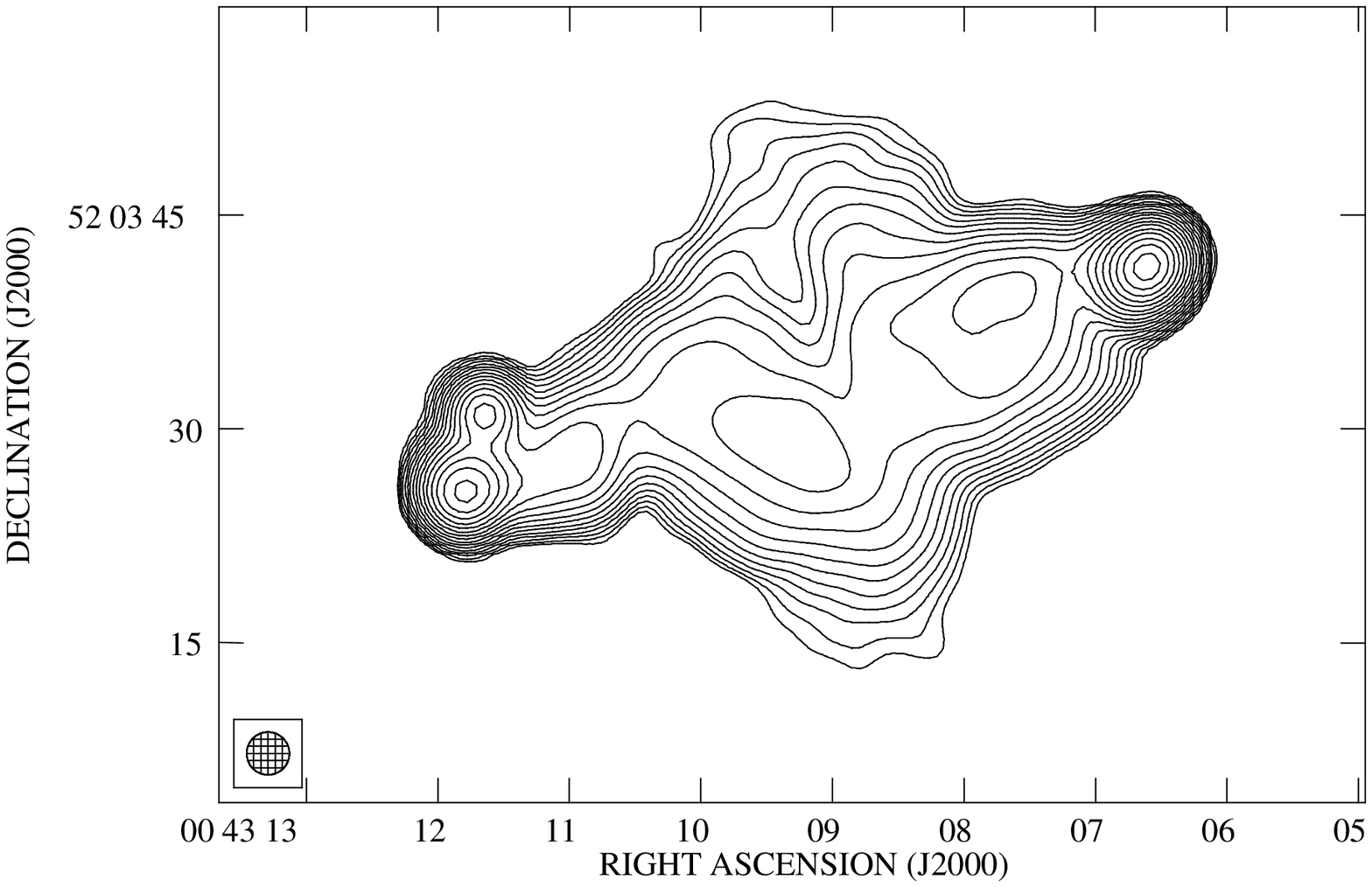}
\begin{center}
1.4 GHz (L band)
\vskip 10pt
\end{center}
\end{minipage}
\begin{minipage}{8cm}
\epsfxsize 8cm
\epsfbox{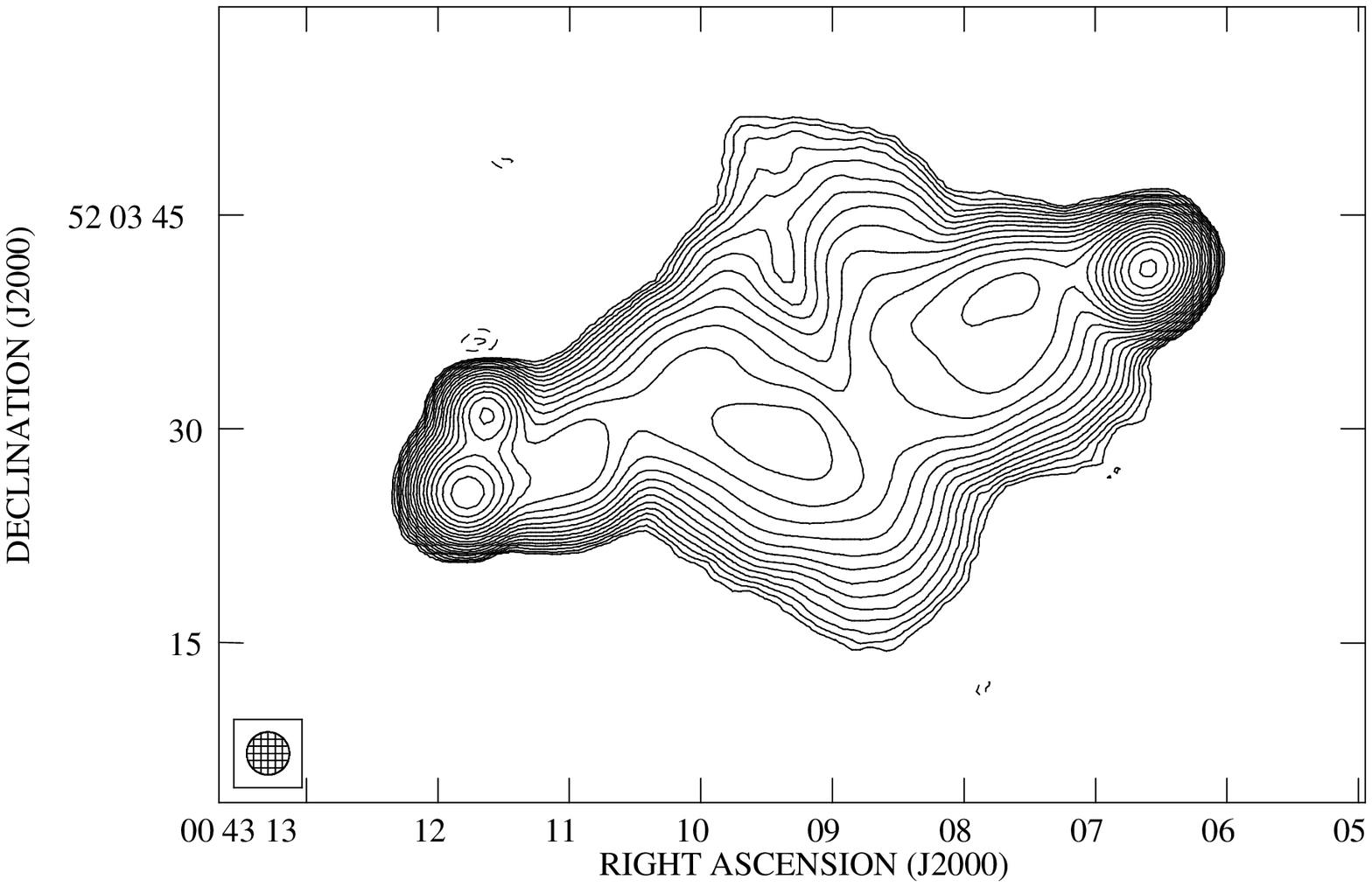}
\begin{center}
4.9 GHz (C band)
\vskip 10pt
\end{center}
\end{minipage}
}
\hbox{
\begin{minipage}{8cm}
\epsfxsize 8cm
\epsfbox{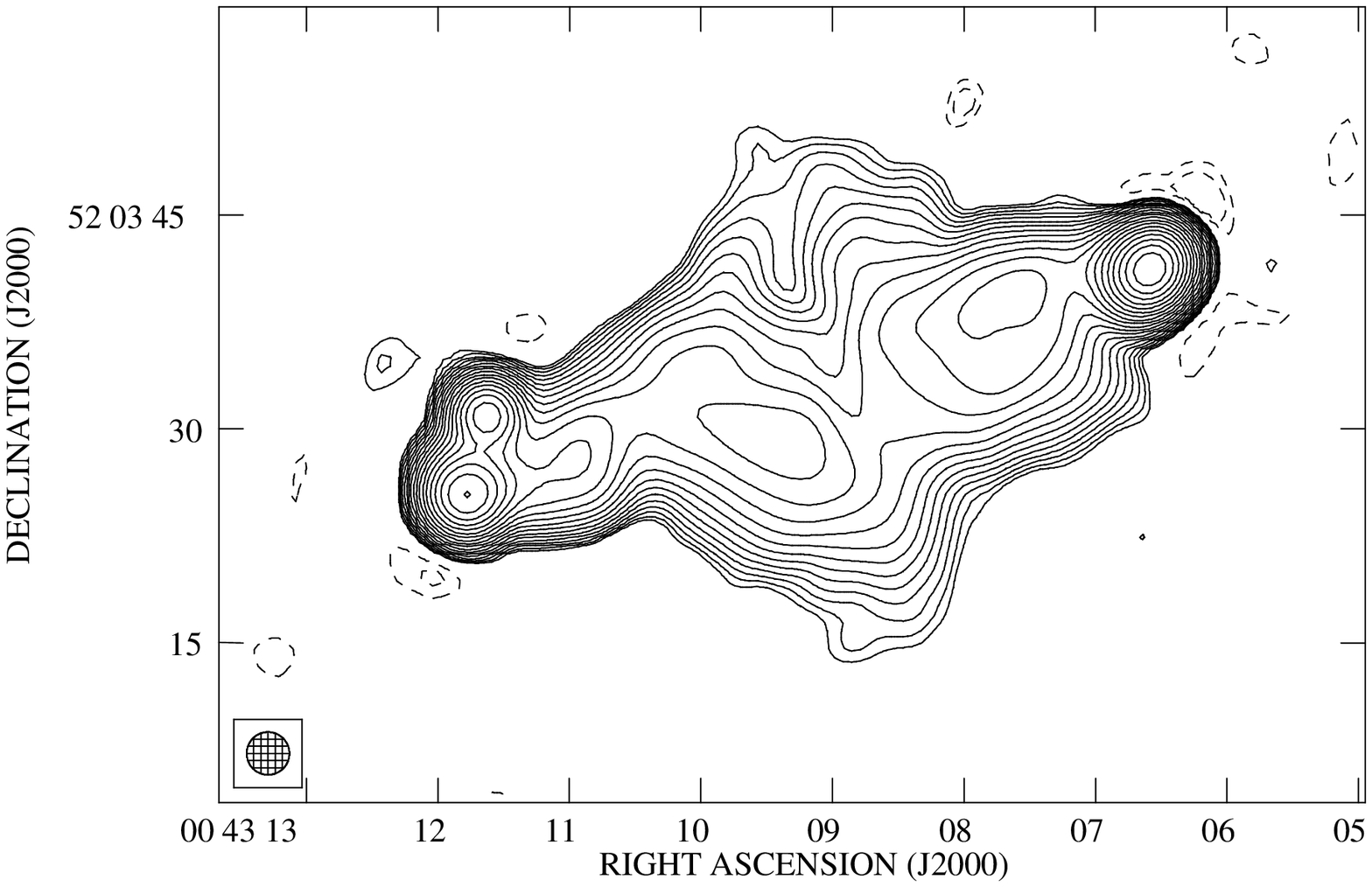}
\begin{center}
8.5 GHz (X band)
\end{center}
\end{minipage}
\begin{minipage}{8cm}
\epsfxsize 8cm
\epsfbox{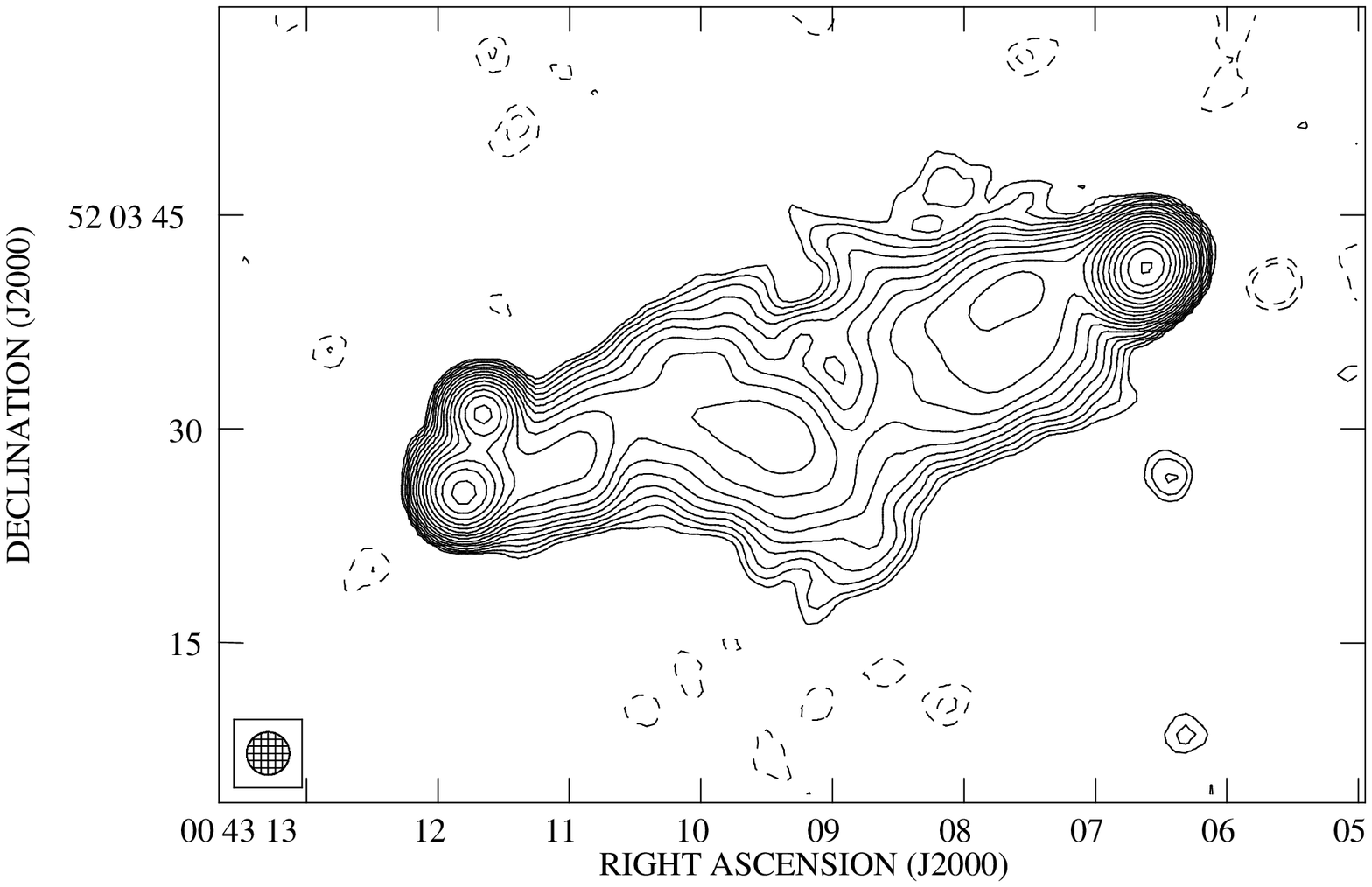}
\begin{center}
14.9 GHz (U band)
\end{center}
\end{minipage}
}
\end{center}
\end{figure*}

To compare our data with observations at longer wavelengths, we
obtained existing VLA data or images at 1.4, 4.9, 8.5 and 15 GHz. The
1.4 GHz image was taken from Leahy, Bridle \& Strom (1998) based on
unpublished observation by R.A. Laing with the VLA A
configuration. The 8.5-GHz data were described by Hardcastle \etal\
(1997) and used the VLA A, B and C configurations. We retrieved 4.9- and
15-GHz snapshot observations taken in 1987-88 from the VLA archive,
using B and C configurations and C and D configurations respectively,
and calibrated and reduced them in the standard manner within AIPS,
using 3C\,286 as the primary flux calibrator. The flux calibration is
estimated to be accurate to within 2 per cent. [In the case of the
D-configuration 15-GHz observations no scan on a suitable calibration
source was made, and we calibrated the amplitude by reference to the
radio galaxy 3C\,123, whose 15-GHz total flux density we know (Looney \&
Hardcastle 2000).] All these datasets have shortest baselines $\sim 2$
k$\lambda$, well matched to the BIMA observations. For comparison with
Fig.\ \ref{bimapix}, in Fig.\ \ref{vlamaps} we show VLA images at
these four frequencies with a resolution of 3 arcsec.

\section{Spectra}

\begin{table*}
\caption{Flux densities from components of 3C\,20's hotspots}
\label{fluxes}\begin{tabular}{llrrrrrr}
\hline
Source&Method&\multicolumn{6}{c}{Flux density (mJy)}\\
component&&1.4 GHz&4.9 GHz&8.5 GHz&15 GHz&85 GHz&231 GHz\\
\hline
W hotspot&Region&$2661 \pm 2$&$1112 \pm 0.4$&$714 \pm 0.2$&$413 \pm
0.4$&$118 \pm 3$&$51.4 \pm 5.4^\dag$\\
&Gaussian&$2515 \pm 2$&$1055 \pm 0.2$&$684 \pm 0.2$&$396 \pm 0.4$&$114
\pm 2$\\
&Compact&--&$250 \pm 5^*$&$179 \pm 3$ &$123 \pm 2^*$&--&--\\
&Extended&$1648 \pm 43$&$650 \pm 7$&$387 \pm 0.7$&$186 \pm 3$&$43 \pm
5$&$11 \pm 6$\\
NE hotspot&Region&$682 \pm 2$&$277.2 \pm 0.3$&$169.0 \pm 0.15$&$103.1 \pm
0.4$&$22.4 \pm 2.2$\\
&Gaussian&$492 \pm 2$&$209.1 \pm 0.3$&$130.7 \pm 0.2$&$81.9 \pm
0.3$&$21.3 \pm 2$&--\\
&Compact&--&$137 \pm 5^*$&$87 \pm 3$&$61 \pm 2^*$&--&--\\
&Extended&$271 \pm 33$&$108 \pm 4$&$55.7 \pm 0.3$&$27 \pm 2$&$0.6 \pm 3.1$\\
SE hotspot&Region&$1257 \pm 2$&$486.4 \pm 0.3$&$289.3 \pm 0.14$&$171.2\pm
0.34$&$36.5 \pm 2.1$\\
&Gaussian&$1280 \pm 2$&$491.9 \pm 0.3$&$292 \pm 0.2$&$166 \pm
0.4$&$40 \pm 3$&--\\
E lobe&Region&$2141 \pm 4.1$&$718.6 \pm 0.7$&$389 \pm 0.3$&$215 \pm
0.2$&$18 \pm 5$\\
\hline
\end{tabular}
\begin{minipage}{14.4cm}
In column 2, `Region' implies that the measurements were made from
direct integration of rectangular regions on 3-arcsec resolution maps,
and the errors are derived directly from the off-source
noise. `Gaussian' implies that the measurements were made by fitting a
Gaussian and background to the same maps, and the errors are the
values returned by the AIPS task {\sc jmfit}. `Compact' implies that
the flux density was determined by Gaussian-fitting to the highest-resolution
radio maps, at resolutions of $\sim 0.2$ arcsec, and the errors are
derived from Gaussian fitting with a variation in the choice of
regions, as described by H94. `Extended' implies that the flux densities are
calculated by subtracting an extrapolation of the power-law spectrum
of the `compact' components from the flux densities measured by
integration, as described in the text; the errors include a
contribution (which dominates at high and low frequencies) from the
calculated uncertainty on the best-fit slope. 4.9- and 15-GHz points
marked with an asterisk are taken from H94; the 231-GHz data point
marked with a dagger is from the IRAM observations of Meisenheimer
\etal\ (1989).
\end{minipage}
\end{table*}

\begin{figure*}
\caption{8.5-GHz radio images of 3C\,20's hotspots (data from
Hardcastle \etal\ 1997) at $0.19 \times 0.16$-arcsec resolution.
Contours are at $0.1 \times (-\protect\sqrt 2, -1, 1,
\protect\sqrt 2, 2, \dots)$ mJy beam$^{-1}$. The eastern hotspot pair is on
the left and the western hotspot on the right.}
\label{radiomap}
\begin{center}
\leavevmode
\hbox
{
\epsfxsize 7.5cm
\epsfbox{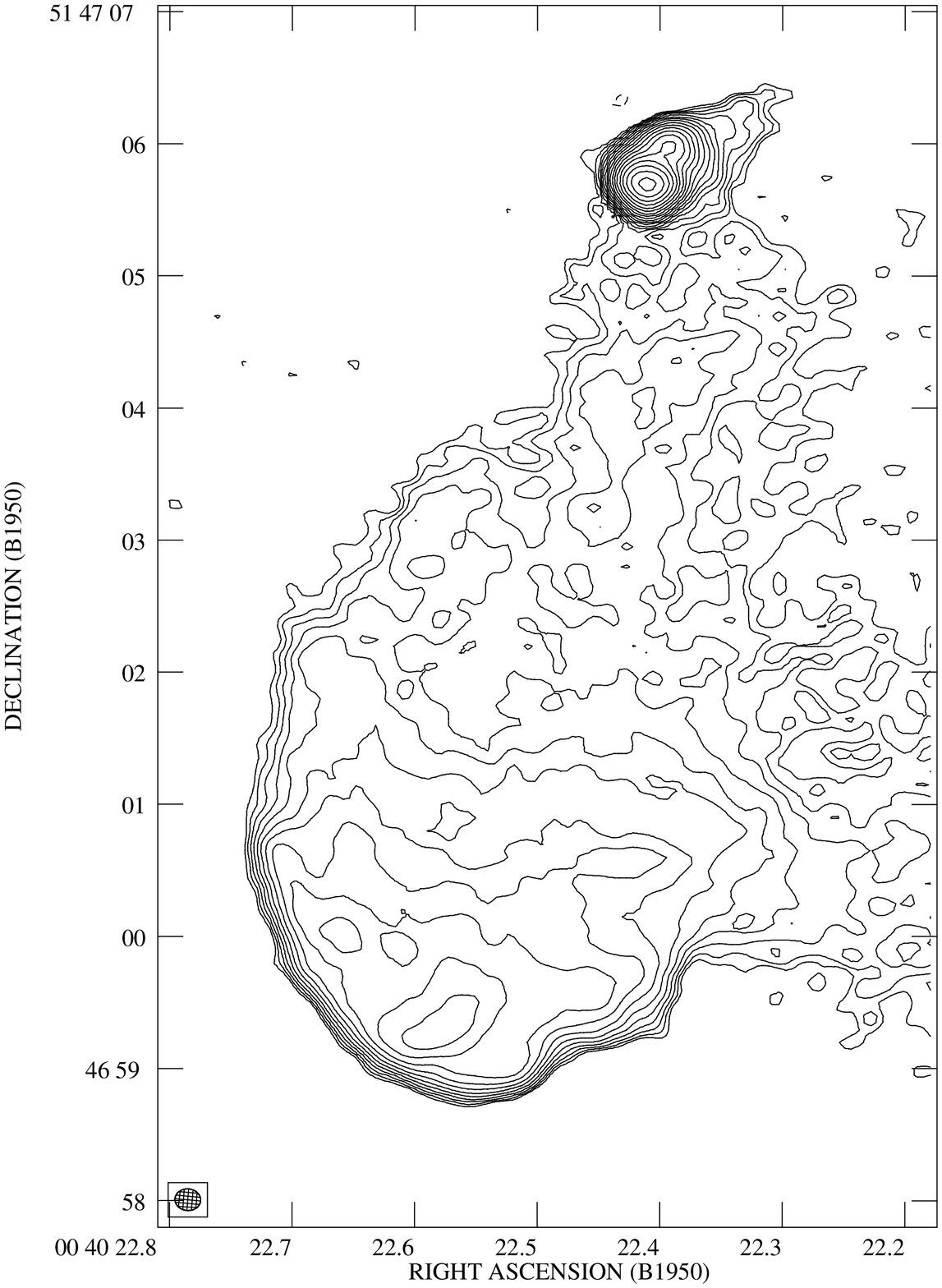}
\epsfxsize 7.5cm
\epsfbox{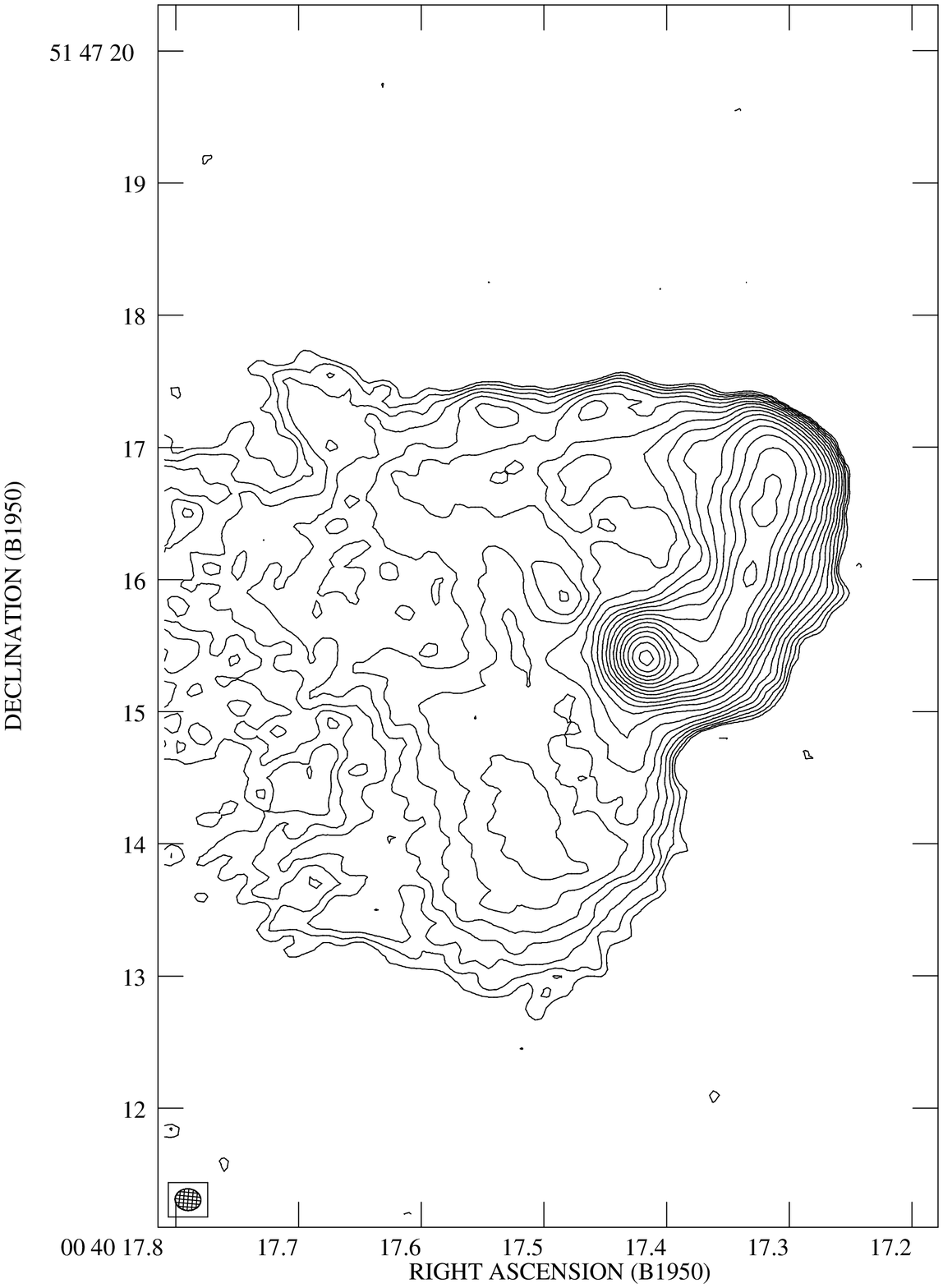}
}
\end{center}
\end{figure*}

\begin{figure}
\caption{Spectral index of 3C\,20 between 1.4 and 8.5 GHz at
1.5-arcsec resolution. Boxes show
the regions for which flux densities were measured.}
\label{spix}
\epsfxsize\linewidth
\epsfbox{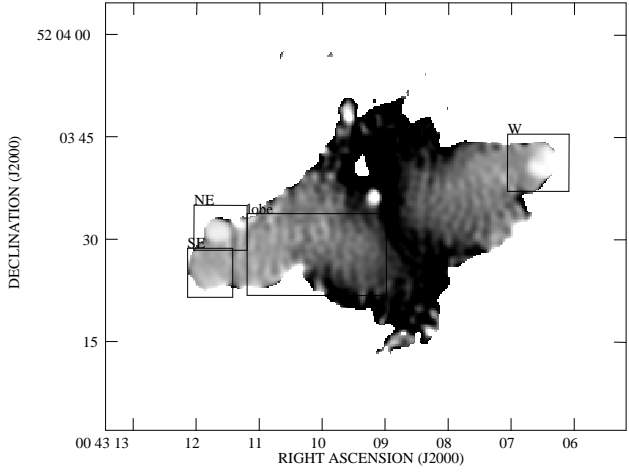}
\end{figure}

The low resolution of the BIMA data compared to the size of the
compact hotspots of 3C\,20 (Fig.\ \ref{radiomap}) means that there is
not a single obviously correct procedure for using these data to
constrain the hotspot spectra. The 3-arcsecond beam certainly contains
more than one spectral component in the case of the compact hotspot in
the eastern lobe (hereafter the `NE hotspot') and the compact pair of
hotspots in the western lobe (the `W hotspot'), because these two
components are surrounded by more or less diffuse emission with a
steeper spectrum at cm wavelengths (Fig.\ \ref{spix}) which may or may
not contribute to the 85-GHz flux density. On the other hand, the secondary
hotspot in the E lobe (the `SE hotspot') might well be spectrally
homogeneous, given its relaxed appearance at high resolution.

We have therefore chosen to make radio measurements of the flux
densities of these components in several different ways. The first and
most obvious involves making radio maps of similar resolution to the
BIMA data (as in Fig.\ \ref{vlamaps}), and then integrating over fixed
regions. In addition to the hotspots, we have measured a flux density
from the E lobe of the source, weakly detected in the BIMA
observations. The regions used are shown on Fig.\
\ref{spix}. Secondly, we can also fit a Gaussian and background to
identical regions centred on the hotspots.  Both integration and
Gaussian fitting will tend to produce values of radio flux density
which include a contribution from the extended emission around the
hotspot.

Thirdly, we can compare the 85-GHz flux densities to the radio flux densities
of the most compact radio components only. This was the approach taken
by Meisenheimer \etal\ (1989) in their study of the W hotspot. The
assumption here is that the most compact component will be
flat-spectrum and will dominate increasingly at higher
frequencies. For this purpose we take flux density values for the
compact hotspots at 4.9 and 15 GHz from H94, and measure values at 8.5
GHz by fitting a Gaussian and baseline to the full-resolution map of
Fig.\ \ref{radiomap}. This technique can be applied only to the W and
NE hotspots. There is some ambiguity in applying it to the NE hotspot,
which at high resolution has double structure (a bright component and
a weak component separated by 0.3 arcsec), as shown in Fig.\
\ref{radiomap}. For consistency with the values of H94 we tabulate the
flux density of the brighter component only; for reference, the integrated
8.5-GHz flux density of the weaker component is about 23 mJy.

Results from these three approaches are tabulated in Table
\ref{fluxes}. The compact components in the NE and W hotspots are
well fitted by power laws with spectral indices of $0.72 \pm 0.04$ and
$0.63 \pm 0.02$ respectively -- these values agree well with the
two-point spectral indices of H94. There is no evidence for a spectral
break at cm wavelengths as there was in the hotspots of 3C\,123
(Looney \& Hardcastle 2000).

In most cases flux densities measured by integrating regions of the
source at 3-arcsec resolution are similar to fluxes derived from
Gaussian fitting, as expected. In the remaining analysis we will only
use the fluxes measured by integration.

\subsection{The eastern double hotspot}

\begin{figure}
\caption{Spectra of the NE and SE hotspots. Solid lines are best-fit
power-law models to the data.}
\label{doublespec}
\epsfxsize \linewidth
\epsfbox{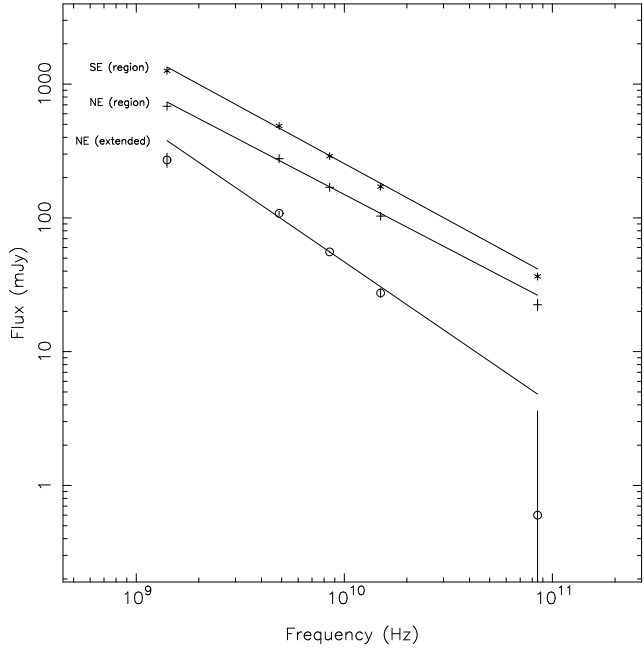}
\end{figure}

It is immediately obvious from Table \ref{fluxes} that the overall
spectral indices of the NE and SE hotspot components are quite
similar; the spectral index over the full frequency range,
$\alpha_{1.4}^{85}$, is $0.83 \pm 0.03$ for the NE hotspot and $0.86
\pm 0.02$ for the SE hotspot. The two spectra are plotted in
Fig.~\ref{doublespec}. There is therefore no immediate evidence for
differential spectral ageing in the more extended SE component. Could
this just be due to contamination of the NE hotspot flux densities by
steep-spectrum extended material included in the integration region?
To test this, we subtracted the best-fit power-law spectrum of the
compact components from the integrated flux densities, assuming that
both compact components of the NE hotspot have the same power-law
spectral index of 0.72. The resulting `extended' flux densities for
the NE hotspot are also tabulated in Table \ref{fluxes}, and it can be
seen that almost all the 85-GHz flux of the hotspot can be accounted
for as an extrapolation of the spectrum of the compact components.
The spectrum of the extended component of the NE hotspot is plotted in
Fig.\ \ref{doublespec}. The integrated spectrum of the NE hotspot can
therefore consistently be modelled as a sum of compact, flat-spectrum
($\alpha = 0.72$) and extended, steep-spectrum ($\alpha \sim 1$)
components, with the expended component possibly cutting off before 85
GHz. But it is clear that the spectrum of the SE component is
consistently flatter than the spectrum of the {\it extended} component
of the NE hotspot, except at low energies where they are comparable.

In detail, the spectrum of the SE hotspot is itself somewhat peculiar;
the spectral index in the range 5--15 GHz is around $0.93 \pm 0.01$,
consistent with H94's estimate of $0.91$ with a scatter of $0.08$, but
it then seems to flatten slightly, with $\alpha_{15}^{85} = 0.89
\pm 0.03$. The inconsistency is only marginal, but it may
indicate that a multi-component model is also necessary to explain the
SE hotspot's spectrum; perhaps the SE hotspot is also a superposition
of flat- and steep-spectrum components. If so, it is puzzling that H94
find a relatively uniform spectral index across the hotspot, a result
confirmed by our own maps of $\alpha_{1.4}^{8.5}$ (Fig.\ \ref{spix});
there is no obvious site for a flat-spectrum region.

What are the implications of these results for models of the double
hotspot? We begin by modelling the spectrum of the compact component
of the NE hotspot. Using an angular size derived from our fits to the
8.5-GHz map (which are consistent with the fits of H94), we can model
the hotspot as a sphere of radius 380 pc. Assuming that the electron
distribution follows a power law with a minimum energy of $5 \times
10^7$ eV, that the electron power-law index is 2.44 (equivalent to the
observed spectral index of 0.72), and that no protons are present in
the hotspot, we obtain an equipartition\footnote{We feel justified in
using equipartition arguments to estimate the field strengths in
hotspots, given the detections of X-ray emission at levels consistent
with inverse-Compton emission at equipartition in the hotspots of Cygnus A
(Harris, Carilli \& Perley 1994), 3C\,295 (Harris \etal\ 2000) and
3C\,123 (Hardcastle \etal\ in prep.).} field strength fit of 36
nT. Using this, we can constrain the high-energy cutoff of the
electron spectrum. The maximum value of the high-energy cutoff
consistent with the upper limits on optical flux density derived by
H94 is $1.6 \times 10^{11}$ eV; the minimum possible high-energy
cutoff if this component is to account for most of the 85-GHz flux
density from the NE hotspot is around $9 \times 10^9$ eV ($3\sigma$
lower limit).

The secondary hotspot's spectrum is not very consistent with an
aged-synchrotron model, because of the flattening of the spectrum
between 15 and 85 GHz. It is poorly fit even with a power-law model
(best-fit $\chi^2 = 14$ with 3 degrees of freedom for $\alpha = 0.84
\pm 0.01$, where the errors include a contribution from the
uncertainty in the flux scales). If we take the slope of this fit as
an indication of the `injection index', the spectral index
corresponding to the low-energy electron energy distribution power-law
index, and we assume standard Jaffe \& Perola (1973, hereafter JP)
ageing in a 6-nT field (the equipartition value), then the 85-GHz data
point allows us to rule out at the 99 per cent confidence level
($\Delta \chi^2 = 6.6$) ages for the SE hotspot greater than $4 \times
10^4$ years; in other words, particle acceleration in the hotspot is
either still going on (in which case a JP spectrum is not appropriate)
or it ceased less than $4 \times 10^4$ years ago. This gives a
disconnection timescale of less than one per cent of the total age of
the source inferred by spectral ageing methods from the spectrum of
the oldest material (e.g.\ Stephens 1987) but, as we pointed out in
our discussion of 3C\,123 (Looney \& Hardcastle 2000) numerical
simulations show that hotspots are transient features on this sort of
timescale. We note that if particle acceleration {\it is} going on in
the hotspot, then this calculation does not tell us anything useful
about the hotspot's age. In this case we would expect a continuous
injection (CI) spectrum rather than a JP spectrum (e.g.\ Pacholczyk
1970), and the position of the break in CI spectra does not bear the
same simple relation to the `age' of the electron population as the
break in JP spectra.

In the next section we use this spectral information to discuss models
for the formation of the double hotspot.

\subsection{Models for the double hotspot}

We can rule out what might be called a trivial dentist's-drill model
in which the jet, having made a single NE-type hotspot at the SE
hotspot position, moves to the NE position and makes another, leaving
the SE hotspot to expand. In this respect 3C\,20 is an extreme example
of the arguments put forward by e.g.\ Valtaoja (1984). The SE hotspot
is not only slightly more luminous, but also very much larger than the
NE hotspot; for the purposes of modelling, we treat it as a hemisphere
of radius 2 arcsec (8 kpc). The `trivial dentist's-drill model' is
then not supported by the energetics of the situation; the
(equipartition) total energy in the SE hotspot is $\sim 10^2$ times
greater than that in the NE hotspot (a point which we return to
below). But it is even more strongly ruled out by expansion losses,
without the need for the assumption of equipartition. To get from a
component comparable to the NE hotspots to the SE hotspot by linear
expansion, the expansion factor would need to be $\sim 20$. Adiabatic
losses coupled with magnetic field conservation (e.g.\ Longair, Ryle
\& Scheuer 1973) would mean that the radio flux density of the
progenitor of the SE hotspot before expansion would have been absurdly
high, a factor $10^6$ higher than its present value; but they would
also mean that the high-energy cutoff in the hotspot would now be $\la
8 \times 10^9$ eV, based on the upper limit on the high-energy cutoff
in the NE hotspot, which, given an equipartition field strength of
$\sim 6$ nT estimated from the observations, is far too low to produce
the observed 85-GHz emission from the SE hotspot. The secondary
component in 3C\,20 cannot have evolved by adiabatic expansion from a
single primary with properties even slightly resembling those of the
NE hotspot, given a simple geometry. The problem is only slightly
alleviated if we consider less simple geometries for the emitting
material in the southern hotspot.

Instead, we can consider a more sophisticated version of the dentist's
drill model, in which the jet termination point moves around in the
region of the SE hotspot before switching to that of the current
primary. In this case, the energy available to the SE hotspot is
potentially that of a large number of components more similar to the
NE hotspot, which reduces the problem of energetics. However, the
adiabatic expansion argument is harder to evade in this way; the
multiple versions of the primary hotspot left in the volume of the SE
hotspot will all expand adiabatically once the jet moves on to the
next one, and so each in turn will become fainter by a large
factor. Clearly this model requires just on energetic grounds that the
jet has dwelt in the region of the SE hotspot at least a hundred times
as long as it has been powering the NE hotspot, and probably much
longer once adiabatic losses are taken into account. For this model to
be viable we therefore have to be seeing the source at a very special
time in its history, and while this is not impossible it is not an
explanation that we wish to adopt while others exist.

The other possible models all require that there is, or was until
recently, long-term energy input into the SE hotspot from a source
other than a `primary-type' jet. The energy input may either be simple
advection of pre-accelerated particles, or it may predominantly be in
the form of kinetic energy of a beam, with particle (re)acceleration
taking place in the SE hotspot. We comment on several possible models
in turn.

\begin{itemize}
\item {\it Disconnected-jet models.} In these models, described by Cox
\etal , the remains of a disconnected jet continue to power the
secondary hotspot after a new primary has started to form. This would
help to explain the low age of the material in the secondary
hotspot. We see two problems with these models in the case of
3C\,20. One is that the relaxed appearance of the hotspot suggests
that there is not currently well-collimated energy input -- a
disconnected jet's impact point should presumably look similar to a
primary hotspot's, since the material of the disconnected jet
presently entering the hotspot does not `know' that it has been cut
off further upstream. The other is that this model does not help to
account for the factor of 100 difference between the energy content of
the SE and NE hotspots; like the dentist's drill model, it requires
that we are seeing the source at a special time.

\item {\it Two-jet models.} We cannot rule out the possibility that
there are two jets, or a split jet, with a powerful but poorly
collimated jet powering the secondary while a weak, well-collimated
jet powers the primary. The only evidence against this is that the
`secondary jet' this would require has never yet been observed in any
radio source.

\item {\it Outflow from the primary}. Laing (1982) was the first to
point out that the SE hotspot is brightest in the region opposite the
compact NE hotspot. That suggests that the secondary may be
powered by direct outflow from the primary, as in the splatter-spot or
redirected-outflow models. The light travel time from the primary to
the secondary hotspot is $\ga 10^5$ years (the uncertainty is due to
the unknown projection angle). So the age constraints derived above
rule out simple advection of particles from the primary, unless they
are transported at very high speeds ($\gamma_{\rm bulk} \ga 2$) or in
a low-loss channel. Instead, it seems most likely that energy is
transported to the SE hotspot largely as bulk kinetic energy, in quite
poorly collimated outflow from the primary, and that particle
acceleration is ongoing there. (The flattening of the secondary
hotspot spectrum between 15 and 85 GHz, and the fact that the
secondary hotspot's spectrum is flatter than that of the extended
component of the primary, may provide some evidence for particle
acceleration.) In this model, the energy content of each hotspot is a
function not just of their history, but of the rate at which energy is
transported {\it out} of the primary and of the efficiency of the
particle acceleration process (there is no reason in principle why the
primary cannot be significantly less efficient than the secondary in
turning the bulk kinetic energy provided by the jet into internal
energy of relativistic particles). The large difference between the
energies of the two hotspots therefore does not need to be explained
in terms of a special observing time in this picture.
\end{itemize}

We conclude that the model with the fewest objections against it is
the outflow model, and therefore join Cox \etal\ (1991) in favouring
that model for this source.

\subsection{The western hotspot}

\begin{figure}
\caption{Spectrum of the W hotspot. Solid lines are best-fit
power-law models to the data.}
\label{whspec}
\epsfxsize \linewidth
\epsfbox{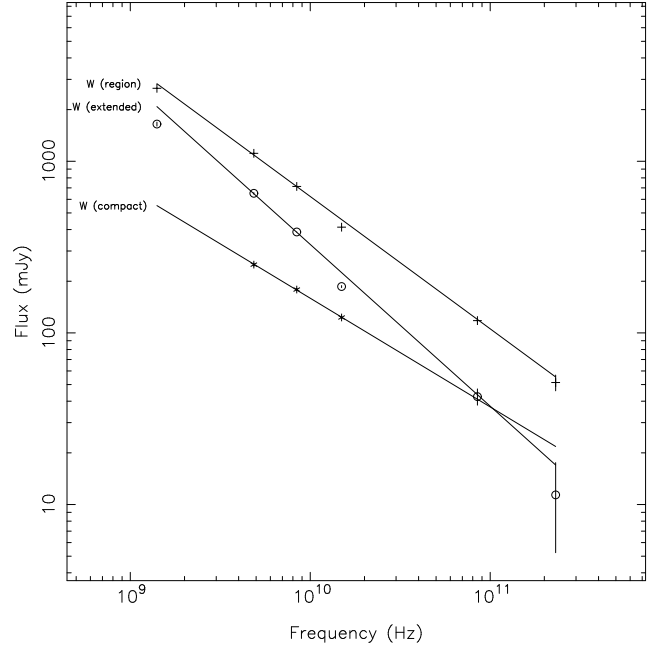}
\end{figure}

The integrated spectrum of the W hotspot (Fig.\ \ref{whspec}) is not
well modelled as a simple power-law. Since the region of the BIMA
detection contains components with a wide range of spectral indices
(Fig.\ \ref{spix}), this is probably not surprising. Table
\ref{fluxes} shows that the most compact component contributes less
than 20 per cent of the total flux density of the integration region even at
15 GHz. The northern `tail' of the compact hotspot (Fig.\
\ref{radiomap}) contributes slightly less, about 150 mJy at 8.5 GHz.
If we assume the tail and hotspot have the same power-law spectral
index of $0.63 \pm 0.02$ (H94 suggest a steeper spectrum for the
northern tail, but this may in part be an effect of a steep-spectrum
background) then we can subtract off an extrapolation of the compact
components to obtain the estimate of the `extended' spectrum tabulated
in Table \ref{fluxes} and plotted in Fig.\ \ref{whspec}. Like the
extended spectrum of the NE hotspot, this is not well fitted either by
a power-law model or by an aged-synchrotron model. But the spectral
shapes of the W and NE extended components are similar, which suggests that
the same sort of physical processes (perhaps related to escape from
the compact hotspots) are at work.

\subsection{The lobes}

Fig.\ \ref{spix} shows that the lobes of 3C\,20 have steep spectra,
which is conventionally attributed to spectral ageing. Some authors
have questioned whether the spectra of lobes can really be described
in this way. Katz-Stone, Rudnick \& Anderson (1993)
show that colour-colour diagrams (where the spectral indices for one
pair of frequencies are plotted against the spectral indices for
another) are useful indicators of whether sources follow any of the
standard ageing models, which correspond to well-defined lines in the
colour-colour plane; they argue that colour-colour diagrams for Cygnus
A are not consistent with any of the standard ageing models. Rudnick
(1999) gives a summary of the problems of spectral ageing.

\begin{figure}
\caption{Colour-colour diagram for 3C\,20; $\alpha_{4.9}^{8.5}$ is
plotted against $\alpha_{1.4}^{4.9}$ for our 3-arcsec resolution
images of the source for each $0.6 \times 0.6$-arcsec pixel with flux
greater than the $3\sigma$ value. The solid line shows the locus of
pure power-law spectra; as discussed by Katz-Stone \etal\ (1993),
lines parallel to but above this line would describe spectra of
constant curvature. The dotted line shows the theoretical track of a
JP spectrum with injection index 0.5 (as predicted by simple particle
acceleration models), and the dashed line the more convincing fit
provided by a JP spectrum with injection index 0.75.}
\label{ccd}
\epsfxsize \linewidth
\epsfbox{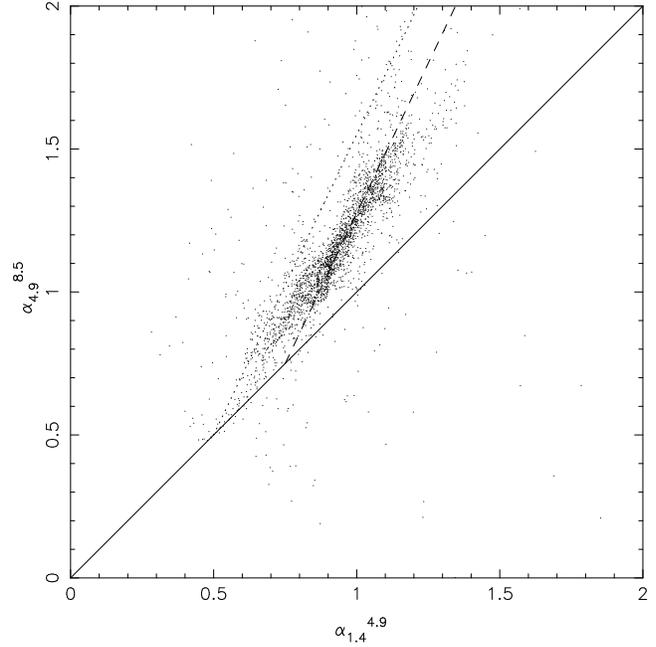}
\end{figure}

In Fig.\ \ref{ccd} we plot a colour-colour diagram for 3C\,20, using the
1.4-, 4.9- and 8.5-GHz VLA data, together with lines showing the
tracks of JP aged synchrotron spectra with injection indices of 0.5
and 0.75. The results are quite similar to those of Katz-Stone
\etal\ The bulk of the source material can be adequately modelled as a JP
aged synchrotron spectrum (with an injection index $\sim 0.75$) but at
flat spectral indices the source deviates from the expected spectrum;
we do not see points which are described by an un-aged power law with
$\alpha = 0.75$. In 3C\,20, unlike Cygnus A, some source material is
well described as a simple power law, but this has $\alpha \sim
0.6$. Breaking down the points in Fig.\ \ref{ccd} by source region, we
find that the points with $\alpha_{1.4}^{4.9} < 0.8$ and
$\alpha_{4.9}^{8.5} < 0.9$ are all in the hotspot regions shown in
Fig.\ \ref{spix}, with the $\alpha \sim 0.6$ points being contributed
by the W hotspot. The points with $\alpha_{1.4}^{4.9} > 0.8$ and
$\alpha_{4.9}^{8.5} > 0.9$, which can be modelled with a JP spectrum,
are essentially all in the lobes of the source. The observation that
there are no points with $\alpha_{1.4}^{4.9} \approx
\alpha_{4.9}^{8.5} \approx 0.75$ thus corresponds to the observation
that spectral ages in the lobes of other FRII sources often do not
extrapolate back to zero age as distance from the hotspots tends to
zero (Alexander 1987, Alexander \& Leahy 1987). Stephens (1987) found
a similar effect in a low-resolution spectral age study of 3C\,20.

The outstanding questions raised by Fig.\ \ref{ccd} are:
\begin{enumerate}
\item Is the apparently adequate description of the lobes with JP aged
synchrotron spectra indicating the real underlying physics, or is it a
coincidence? If the lobes really are simply aged material,
\item why is the injection index of the lobes apparently different
from the (flatter) injection indices in the compact hotspots, and
\item why do we see no source material which has a simple power law
spectrum with spectral index equal to the injection index;
equivalently, why is there no zero-age material with the injection
index which appears to characterize the lobes?
\end{enumerate}

Since the hotspots are transient features, question (ii) can be
avoided, if necessary, by noting that the injection index of the
present-day hotspots need not be the same as the injection indices of
the hotspots present at the time that the lobe material was
accelerated. Alexander (1987) suggests that the answer to question
(iii) may be that the effects of losses in the comparatively high
field of the hotspot are exaggerated by adiabatic expansion, which for
linear expansion by a factor $\Delta$ moves the break frequency down
by a factor $\Delta^4$ if magnetic flux is conserved. For $\Delta \sim
2$ as suggested by Alexander in the case of 3C\,234, this would imply
that we should start to see ageing in the mm region of the hotspot
spectra, for which there is no convincing evidence in the case of
3C\,20; however, because of the low resolution of our BIMA
observations, we cannot really rule out the presence of
steeper-spectrum regions around the flat-spectrum hotspots.

\begin{figure}
\caption{The spectrum of the E lobe region of 3C\,20, with a best-fit
JP aged synchrotron spectrum with injection index 0.8 (solid line).}
\label{lobespec}
\epsfxsize \linewidth
\epsfbox{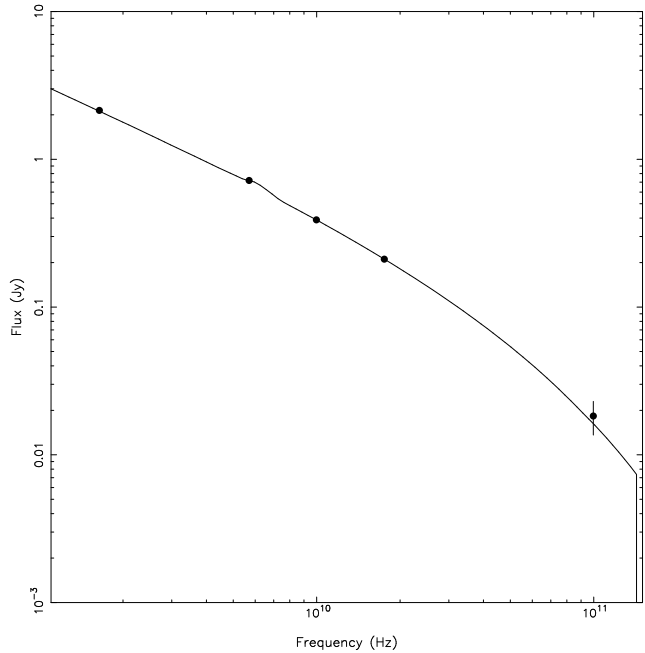}
\end{figure}

An answer to question (i) is beyond the scope of this paper, but we
note that the observation of 85-GHz emission from the lobe region is
very consistent with the predictions of a JP aged model (Fig.\
\ref{lobespec}); if the similarity to a JP spectrum is a coincidence,
it is one which extends over a wide range of frequencies.

\section{Conclusions}

BIMA observations of the double hotspot of 3C\,20 resolve the two
components of the hotspot and, together with archival VLA data, show
that their radio spectra are very similar over the full range studied;
3C\,20 in this respect is similar to other double hotspot sources like
3C\,405 and 3C\,123 previously studied with BIMA. The detailed
spectrum of the secondary hotspot in 3C\,20 provides some evidence
that particle acceleration is continuing, and this, together with the
relaxed and edge-brightened appearance of the secondary in
high-resolution radio maps, causes us to conclude that the most likely
model for this source is one in which the secondary is generated by
direct outflow from the primary hotspot; this outflow can be
relatively slow and poorly collimated compared to the jet.

Both the eastern and western compact (primary) hotspots are bright
sources at 85 GHz, and it seems likely that there is some emission at
this frequency from the regions around them as well. Our radio
observations agree with H94's statement that the radio spectrum of the
eastern primary is slightly but significantly steeper than that of the
W primary, but the 85-GHz observations are consistent with the primary
hotspots being otherwise similar at high frequencies; in fact, our
estimated upper limit on the high-energy cutoff in the compact
component NE hotspot is exactly consistent with the high-energy cutoff
required to produce the observed optical emission from the compact
component of the W hotspot, if the flatter injection index and
slightly higher equipartition field strength in the W hotspot are
taken into account. Further observations at frequencies intermediate
between 85 GHz and optical will be required to establish whether the
primary hotspots, so similar in size and energy content, have
different high-energy cutoffs in their electron spectra, and to
establish a difference between the NE and SE hotspots.

Our VLA and BIMA data show that 3C\,20's lobes are quite well
described by a simple JP ageing model, but that, as with other
sources, it is difficult to understand either why the injection
indices in the lobes and hotspots are different or why no material
with a pure injection-index spectrum is seen.

\section*{ACKNOWLEDGEMENTS}
We thank an anonymous referee for comments which enabled us to improve
the paper significantly. We are grateful to Rick Perley for allowing
us to use the archival 5- and 15-GHz VLA data discussed in the text,
and the VLA data analysts for help in extracting it from the
archive. The National Radio Astronomy Observatory VLA is a facility of
the U.S. National Science Foundation operated under cooperative
agreement by Associated Universities, Inc.

\clearpage

\end{document}